\documentclass{emulateapj}
\usepackage{color}

\usepackage{mathrsfs}
\usepackage{graphicx}
\usepackage{cases}
\usepackage{txfonts}
\usepackage{epsfig}
\usepackage{multirow}

\def\prd{Phys. Rev. D}

\def\prl{Phys. Rev. Lett.~}

\def\apj{Astrophys. J.~}
\def\apjl{Astrophys. J. Lett.~}
\def\araa{Annu. Rev. Astron. Astrophys.~}
\def\mnras{Mon. Not. Roy. Astron. Soc.~}
\def\apjs{Astrophys. J. Suppl. Ser.~}

\def \nat{Nature}

\def\msun{M_{\odot}}

\def\be{\begin{equation}}
\def\ee{\end{equation}}

\def\bea{\begin{eqnarray}}
\def\eea{\end{eqnarray}}

\begin{document}

\title{On using inspiralling supermassive binary black holes in the PTA frequency band as standard sirens to constrain dark energy}

\author{Changshuo Yan$^{1,2\,\dagger}$, Wen Zhao$^{3,4\,\star}$, and Youjun Lu$^{1,2\,\ddagger}$}
\affil{$^{1}$\,CAS Key Laboratory for Computational Astrophysics, National Astronomical Observatories, Chinese Academy of
Sciences, Beijing, 100101, China; $^\dagger$\,yancs@nao.cas.cn, $^\ddagger$\,luyj@nao.cas.cn\\
$^{2}$\,School of Astronomy and Space Science University of Chinese Academy of Sciences, Beijing, 100049, China\\
$^{3}$\,CAS Key Laboratory for Researches in Galaxies and Cosmology, Department of Astronomy, University of Science and Technology of China, Chinese Academy of Sciences, Hefei, Anhui 230026, China; $^\star$\,wzhao7@ustc.edu.cn  \\
$^{4}$\,School of Astronomy and Space Science, University of Science and Technology of China, Hefei, 230026, China}

\begin{abstract}
{
Supermassive binary black holes (SMBBHs) in galactic centers may radiate gravitational wave (GW) in the nano-Hertz frequency band, which are expected to be detected by pulsar timing arrays (PTAs) in the near future. GW signals from individual SMBBHs at cosmic distances, if detected by PTAs, are potentially powerful standard sirens that can be used to independently measure  distances and thus put constraints on cosmological parameters. In this paper, we investigate the constraint that may be obtained on the equation of state ($w$) of dark energy by using those SMBBHs, expected to be detected by the PTAs in the Square Kilometre Array (SKA) era. By considering both the currently available SMBBH candidates and mock SMBBHs in the universe resulting from a simple galaxy major merger model, we find that $\sim 200$ to $3000$ SMBBHs with chirp mass $>10^9M_\odot$ are expected to be detected with signal-to-noise ratio $>10$ by SKA-PTA with conservative and optimistic settings and they can be used to put a constraint on $w$ to an uncertainty of $\Delta w\sim 0.02-0.1$.  If further information on the mass and mass ratio of those SMBBHs can be provided by electromagnetic observations (e.g., chirp mass uncertainty $\lesssim 50\%$), the constraint may be further improved to $\lesssim 0.01$ level, as many more SMBBHs will be detected by SKA-PTA with relatively better distance measurements and can be used as the standard sirens.
}
\end{abstract}

\keywords{Active galactic nuclei (16), Black hole physics (159),  Cosmological parameters (339), Dark energy (351), Fisher's Information (1922), Gravitational waves (678), Gravitational wave astronomy (675), Pulsar timing method (1305), Supermassive black holes (1663) }

\section{Introduction }
\label{intro}

It is crucial to accurately measure the cosmological parameters for understanding the dynamical evolution of the universe and the nature of dark matter and dark energy. Numerous methods have been developed to achieve this goal. However, current available measurements obtained by using different methods may have significant discrepancy, e.g., the $4.4$-$\sigma$ discrepancy between the Hubble constant ($H_0$) inferred from the Planck cosmic microwave background (CMB) data and that obtained from SN Ia standard candles \citep[see][]{Planck2018, Riess19}. This ``Hubble tension'' may be an indicator of new physics beyond the standard Lambda cold dark matter ($\Lambda$CDM) cosmology or unknown systemic biases in those current methods. Therefore, it is important to propose and apply other (new) method(s) to independently measure cosmological parameters and compare them with those traditional methods for improving the measurement accuracy of the cosmological parameters.

Gravitational wave (GW) from a compact binary coalescence (CBC) provides a new type of ``standard siren'' to independently probe cosmological parameters, if its redshift can be measured \citep{Schutz86, ChernoffandFinn93, Finn96}. The first multi-messenger detection of a double neutron star (DNS) merger, GW170817 \citep{Abbott2017a, Abbott2017b}, has enabled the first standard siren measurement of $H_0$ \citep{Abbott2017c, Hotokezaka19}, and demonstrated the great potential of this un-biased method \citep[e.g.,][]{Chen2018, Zhao2011, Zhao2018}. However, only mergers of DNSs and black hole-neutron star binaries are expected to have significant electromagnetic (EM) signals (though weak), with which their redshift information can be obtained. Large fraction of the sources detected by ground-based GW detectors would be mergers of stellar-mass binary black holes (sBBHs),
which may not be accompanied with significantly bright EM counterparts as current searches for their EM counterparts all returned null result \citep[e.g.,][]{2019ApJ...886...73N}.
This may significantly limit the distance and number of GW sources that can be used as the standard sirens, and thus limit the power of this method to measure the cosmological parameters and constrain the nature of dark matter and dark energy.

Inspiralling of supermassive binary black hole (SMBBHs; with mass $\gtrsim 10^8 M_\odot$) in galactic centers is important GW sources at $10^{-9}-10^{-6}$\,Hz, which are long anticipated to be detected by pulsar timing arrays \citep[PTAs;][]{RR95, JB03, WL03, Sesana13, Ravi14, Perera18}.
Most traditional PTAs studies focus on the detection of the stochastic GW background from numerous cosmic SMBBHs \citep[e.g.,][]{Jenet06, LS2015, Arzoumanian16, Shannon15, Desvignes16, Reardon16, Sesana18, Arzoumanian18, Arzoumanian18b, Perera19}, while recent PTAs studies begin to investigate the detectability of individual SMBBHs \citep{Sesana09, SV2010, FL2010, Lee11, BS2012, Ellis12, Wang14, Ravi14, Arzoumanian14, Zhu15, Madison16, WM2017, Aggarwal19}  and find that the loudest SMBBHs may have rather high signal-to-noise (SNR) to be detected by future PTAs, such as the Square Kilometre Array (SKA) PTA  \citep[e.g.,][]{Ravi15, Rosado15}. 

It is possible that future ``PTAs detected SMBBHs'' can be also taken as the standard sirens to probe cosmology. Different from sBBHs, many SMBBHs may have EM counterparts and thus can be detected by EM waves with redshift measurements. Simulations have also shown that the physical parameters of SMBBH systems (including the luminosity distance) detected by future PTA(s) can be extracted with high accuracy \citep[e.g.,][]{Wang14, Zhu15, WM2017}. Therefore, conceptually it is undoubtable that those PTAs SMBBHs can be used as standard sirens.
However, whether these ``PTAs detected SMBBHs'' can provide sufficiently interesting measurements on the cosmological parameters depends on their foreseeable SNRs and number distribution as a function of redshift.  In this paper, we will investigate the potential of using future ``PTAs detected SMBBHs'' as standard sirens to probe cosmological parameters, especially on constraining the nature of dark energy.

This paper is organized as follows. In Section~\ref{sec:method}, we introduce a method of using Fisher information matrix to analyze the GW signal from SMBBHs and determine measurement errors of various physical parameters involved in. We show how to obtain constraints on the dark energy by using PTAs SMBBHs in Section~\ref{sec:darkenergy}. We illustrate the effects of different physical parameters of the GW sources on the GW detection and the errors of luminosity distance measurements in Section~\ref{sec:effects}. In Sections~\ref{sec:mbbhs}, we apply the above method to the currently available SMBBH candidates from astronomical observations and the mock SMBBH samples obtained from a simple model, respectively, and predict the robustness of the constraints that can be obtained from PTAs SMBBHs. Conclusions and discussions are given in Section~\ref{sec:conclusion}.

Throughout the paper, we adopt a flat $\Lambda$CDM cosmology with $\Omega_{\rm m} = 0.3$, $\Omega_\Lambda$=0.7, $h=0.7$, where $h=H_0/100{\rm km s^{-1} Mpc^{-1}}$ \citep[e.g., cf.][]{2016A&A...594A..13P}.


\section{GW signal and analysis method}
\label{sec:method}

A PTA data set is the time of arrivals (ToAs) for pulses from millisecond stable pulsars (MSPs) monitored over a decade or longer with typical cadence of bi-weekly to monthly
\citep{Desvignes16, Reardon16, Arzoumanian18, Perera19}.
The ToA data encodes the information of the MSPs' rotation, the dispersion due to the ionized interstellar medium, and also the binary behavior of the MSP if it is in a binary system, which can be well revealed by standard models.
The GW effects can be seen in the ToA residuals by removing the model-predicted TOAs from the observational TOA data, and various noise processes can be constrained and included in the timing model \citep{Lentati16}. 
The root mean square (RMS) of these time residuals reflects the stability of the pulsar and the quality of the timing data that can be used to measure or constrain the GW signal(s).

Consider a single GW source coming from a direction $\hat{\Omega}$, its induced pulsar timing residuals measured at time $t$ on the Earth can be written as \citep[e.g., see][]{Zhu15}
\be
s(t,\hat{\Omega})=F^{+}(\hat{\Omega})\Delta A_+(t)+F^{\times}(\hat{\Omega})\Delta A_{\times}(t),
\label{s-definition}
\ee
where $F^{+}(\hat{\Omega})$ and $F^{\times}(\hat{\Omega})$ are the antenna pattern functions as given by \citep{Wahlquist1987}:
\begin{eqnarray}
F^{+}(\hat{\Omega})&=&\frac{1}{4(1-\cos\theta)}\left\{(1+\sin^2\delta)\cos^2\delta_{\rm p}\cos[2(\alpha-\alpha_{\rm p})] \right.\nonumber\\
&& \left. -\sin2\delta\sin2\delta_{\rm p}\cos(\alpha-\alpha_{\rm p})
+\cos^2\delta(2-3\cos^2\delta_{\rm p})\right\}, \nonumber \\
F^{\times}(\hat{\Omega})&=&\frac{1}{2(1-\cos\theta)}\left\{\cos\delta\sin2\delta_{\rm p}\sin(\alpha-\alpha_{\rm p}) \right. \nonumber\\
&&\left. -\sin\delta\cos^2\delta_{\rm p}\sin[2(\alpha-\alpha_{\rm p})]\right\}.
\end{eqnarray}
Here $(\alpha, \delta)$ or $(\alpha_{\rm p}, \delta_{\rm p})$ are the right ascension and declination of the GW source or pulsar, and $\theta$ is the opening angle between the GW source and pulsar with respect to the observe
\be
\cos\theta=\cos\delta\cos\delta_{\rm p}\cos(\alpha-\alpha_{\rm p})+\sin\delta\sin\delta_{\rm p}.
\ee
In Equation~(\ref{s-definition}),
$\Delta A_{\{+,\times\}}(t)=A_{\{+,\times\}}(t)-A_{\{+,\times\}}(t_{\rm p})$,
where $t_{\rm p} = t-d_{\rm p}(1-\cos\theta)/c$ is the time at which the GW passes the MSP
with $d_{\rm p}$ representing the pulsar distance, and $A_{\{+,\times\}}(t)$ and $A_{\{+,\times\}}(t_{\rm p})$ contribute to the Earth term and pulsar term, respectively, for which the specific functional forms depend on the type of sources being search for.
For cases considered in the present paper, we assume evolving SMBBHs and thus the frequency of the Earth-term and pulsar-term are not exactly the same, though the difference is tiny for most cases.
For SMBBHs on circular orbits, we have
\begin{eqnarray}
A_{+}(t)&=&\frac{h_0(t)}{2\pi f(t)}\left\{(1+\cos^2\iota)\cos2\psi\sin[\phi(t)\right.\nonumber\\
&&\left.+\phi_0]+2\cos\iota\sin2\psi\cos[\phi(t)+\phi_0]\right \}, \label{A+} \\
A_{\times}(t)&=&\frac{h_0(t)}{2\pi f(t)}\left\{(1+\cos^2\iota)\sin2\psi\sin[\phi(t) \right. \nonumber\\
&&\left. +\phi_0]-2\cos\iota\cos2\psi\cos[\phi(t)+\phi_0]\right \}.
\label{Across}
\end{eqnarray}
Here $\iota$ is the angle between the normal of the binary orbital plane and the line of sight, $\psi$ is the GW polarization angle, $\phi_0$ is a phase constant, and $h_0$ is the intrinsic GW strain amplitude defined as
\be
h_0=2\frac{(GM^z_{\rm c})^{5/3}}{c^4}\frac{(\pi f)^{2/3}}{d_{\rm L}}=2\frac{(GM_{\rm c})^{5/3}}{c^4}\frac{(\pi f_{\rm r})^{2/3}}{d_{\rm c}},
\ee
with $d_{\rm L}$ and $d_{\rm c}$ representing the luminosity and comoving distance to the source, respectively, $M_{\rm c}=M_{\bullet,1}^{3/5} M_{\bullet,2}^{3/5}(M_{\bullet,1}+M_{\bullet,2})^{-1/5}$ representing the binary chirp mass, $M_{\bullet,1}$ and $M_{\bullet,2}$ ($\leq M_{\bullet,1}$) representing the SMBBH component masses. Noted here that only the redshifted chirp mass $M_{\rm c}^z=M_{\rm c}(1+z)$, but not the chirp mass, is directly measurable from GW data; likewise, the rest-frame frequency $f_{\rm r}$ is related to the observed frequency $f$ by $f=f_{\rm r}/(1+z)$ \citep{Maggiore}. This is the reason that independent measurements of the redshifts of GW sources are required for GW cosmology studies, in order to break such a degeneracy between GW measured mass and redshift.

In the quadrupole approximation, the GW phase and frequency that appear in Equations\,(\ref{A+}) and (\ref{Across}) are given by
\be
f(t)=\left[f_0^{-8/3}-\frac{256}{5}\pi^{8/3}\left(\frac{GM^z_{\rm c}}{c^3}\right)^{5/3}t\right]^{-3/8},
\ee
\be
\phi(t)=\frac{1}{16}\left(\frac{GM^z_{\rm c}}{c^3}\right)^{-5/3}\left\{(\pi f_0)^{-5/3}-[\pi f(t)]^{-5/3}\right\},
\ee
where $f_0$ is the observed GW frequency at the time of the first observation \citep[e.g.,][]{Thorne1987}.

We define the SNR of GW signal detected by a PTAs with $N_{\rm p}$ MSPs as
\be
\rho^2=\sum_{j=1}^{N_{\rm p}}\sum_{i=1}^{N}\left[\frac{s_j(t_i)}{\sigma_{t,j}}\right]^2,
\label{eq:SNR}
\ee
where $N$ is the total number of data points for each MSP, $s_j(t_i)$ is the timing residuals of $j$-th MSP at time $t_i$ (see Eq.~(\ref{s-definition})), and $\sigma_{t,j}$ is the RMS of timing noises of the $j$-th MSP.
In this paper, we adopt the Fisher information matrix for parameter estimations. In the case of a network including $N_{\rm p}$ independent MSPs, the Fisher matrix is
\be
\Gamma_{ab}=\sum_{j=1}^{N_p}\sum_{i=1}^{N}\frac{\partial{s(t_i)}}{\sigma_{t,j}\partial{p_a}}\frac{\partial{s(t_i)}}{\sigma_{t,j}\partial{p_b}},
\ee
where $p_a$ and $p_b$ denote the free parameters to be estimated.

For each given GW source, the response of the pulsar network depends on $N_{\rm p}+8$ system parameters, including those of the GW source (i.e., $M_{\rm c}$, $\alpha$, $\delta$, $\iota$, $\psi$, $\phi_0$, $f_0$, $d_{\rm L}$) and distances of MSPs $d_{{\rm p},j}$ $(j=1, 2, \cdots, N_{\rm p})$.
Prior information can be included as $\Gamma_{ab} \rightarrow \Gamma_{ab}-\left< \frac{\partial^2{\ln} P(p_i)}{\partial p_a\partial p_b}\right>$, where $P(p_i)$ is the prior distribution of the parameter $p_i$ \citep[e.g. see][]{Albrecht09}. For the inclination angle $\iota$ we need to consider its prior distribution. As the disk direction is randomly distribute in $4\pi$ solid angel, so the $P(\iota)\propto {\rm sin}(\iota)$, then we will have $\Gamma_{ii} \rightarrow \Gamma_{ii}+\frac{1}{{\rm sin}^2\iota}$ where $p_i=\iota$.
If the GW sources can be identified electromagnetically, then the sky location, redshift, and even the SMBBH mass can be obtained, which may add some additional information into the Fisher matrix.
In this case, a Gaussian prior with width $\sigma_i$ may be placed on the $i^{th}$ parameter, with additional EM information, by adding to the appropriate diagonal element of the Fisher matrix: $\Gamma_{ab} \rightarrow \Gamma_{ab}+\delta_{ai}\delta_{bi}/\sigma_i^2$.
The Fisher matrix is commonly used in many fields to estimate errors in the measured parameters by the expression $\langle\delta p_a\delta p_b\rangle=(\Gamma^{-1})_{ab}$. Once the Fisher matrix $\Gamma_{ab}$ is calculated, the error in measuring the parameter $p_a$ can then be estimated as $\Delta p_a=(\Gamma^{-1})_{aa}^{1/2}$.
%
%

\section{Constraining dark energy}
\label{sec:darkenergy}

In a flat Universe, the luminosity distance can be written as
\be
d_{\rm L}=(1+z)\int_0^z\frac{dz'}{H(z')},
\ee
where $H(z)$ is the Hubble parameter. Given a form of dark energy with density parameter $\Omega_{\rm de}$ and a (constant) equation-of-state (EoS) index $w$, one has
\be
H(z)=H_0\left[\Omega_{\rm m}(1+z)^3+\Omega_{\rm de}(1+z)^{3(1+w)}\right]^{1/2}.
\ee

Similar to \citet{0810.5727}, we are interested in getting a rough sense of the level of accuracy we can expect in extracting the EoS index of dark energy $w$. From the expression of $d_{\rm L}$, it seems possible that one can constrain the full parameter set $(H_0,\Omega_{\rm m},\Omega_{\rm de}, w)$ together by the GW data alone, as long as the reshifts of GW sources are known. Unfortunately, in the previous work \citet{Zhao2011}, we found this globe constraints cannot be realized, due to the strong degeneracy between the background parameters $(H_0,\Omega_{\rm m},\Omega_{\rm de})$ and the dark energy EoS $w$. The same problem also happens in other methods for dark energy detection (e.g., SN Ia and BAO methods). A general way to break this degeneracy is to combine the result with the CMB data, which are sensitive to the background parameters $(H_0,\Omega_{\rm m},\Omega_{\rm de})$, and provide the necessary complementary to the GW data. It has also been discovered in \citet{Zhao2011} that, taking the CMB observation as a prior is nearly equivalent to treat the parameters $(H_0,\Omega_{\rm m},\Omega_{\rm de})$ as known in data analysis. Thus, we use the GW data to constrain the parameters $w$ only. For a single GW source, the error on $w$ can be estimated as
\be
\Delta w=d_{\rm L} \left|\frac{\partial d_{\rm L}}{\partial w}\right|^{-1} \frac{\sigma_{d_{\rm L}}}{d_{\rm L}}.
\ee
Note that, the uncertainty $\sigma_{d_{\rm L}}=\sqrt{(\Delta d_{\rm L})^2+(\tilde\Delta d_{\rm L})^2}$, where $\Delta d_{\rm L}$ is calculated by the Fisher matrix analysis as above, and
\be
\tilde\Delta d_{\rm L}=d_{\rm L}\times 0.066\left(\frac{1-(1+z)^{-0.25}}{0.25}\right)^{1.8},
\ee
which accounts for the uncertainty caused by weak lensing \citep{Hirata2010}.

If we consider a set of GW sources, and $\Delta w_i$ denotes the result of $\Delta w$ derived from the $i$-th GW source. Then the combined constraint becomes
\be
\Delta w= \frac{1.0}{\sqrt{\sum_{i}(\Delta w_i)^{-2}}}.
\label{eq:dw}
\ee

\section{Effects of physical parameters of the GW sources}
\label{sec:effects}

\begin{figure}
\begin{center}
\setlength{\abovecaptionskip}{0.cm}
\setlength{\belowcaptionskip}{-0.cm}
\centerline{
\includegraphics[scale=0.7,angle=-90]{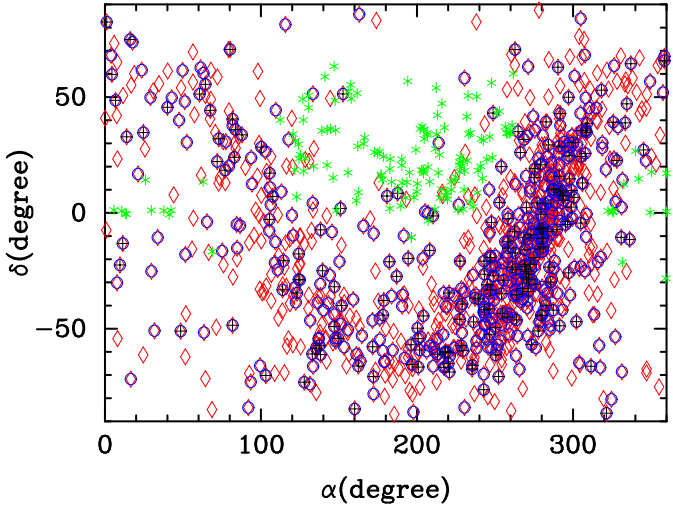}
}
\end{center}
\caption{Positions of the selected $1026$\,MSPs constituting the simulated SKA-PTA and current available SMBBH candidates on the sky. Red diamonds show all of those $1026$\,MSPs, while blue circles and black pluses ($+$) show $500$ and $200$ out of them, respectively. These three different MSP samples are all used in the paper. Green stars show the SMBBH candidates obtained from observations.
}
\label{fig:f1}
\end{figure}

To figure out the effects of the physical parameters of an SMBBH on the GW detection SNR and the distance measurement, we construct an SKA era PTA by using the simulated pulsar catalog in \citet{SKApulsar}. In principle, the choice of MSPs depends on their potential timing accuracy, which mainly depends on the stability of the rotation of MSP itself, and the accuracy of TOA we detect. The neighboring MSPs may have a higher flux, and the integrated pulse profile has a higher SNR and a more accurate timing. In addition, the impact of dispersion and other effects is small. For these reasons, similar to \citet{WM2017}, in this paper we select $1026$\,MSPs within $3$\,kpc from the Earth for the analysis. Figure~\ref{fig:f1} shows the localization of those simulated MSPs. With this assumed SKA-PTA, we generate the data realizations by adopting an uniform cadence, for simplicity, either $1$\,week or $2$\,weeks,while the typical cadence of current PTAs are bi-weekly to monthly.
 The span of the simulated timing residuals is $10$\,years. RMS of timing noises are assumed to be $20$\,ns, $50$\,ns, and $100$\,ns for each MSP, respectively, here we assumed three different values for each simulated pulsar to investigate how the timing precision of pulsars affect the results.

\begin{figure}
\begin{center}
\setlength{\abovecaptionskip}{0.cm}
\setlength{\belowcaptionskip}{-0.cm}
\centerline{\includegraphics[scale=0.32,angle=-90]{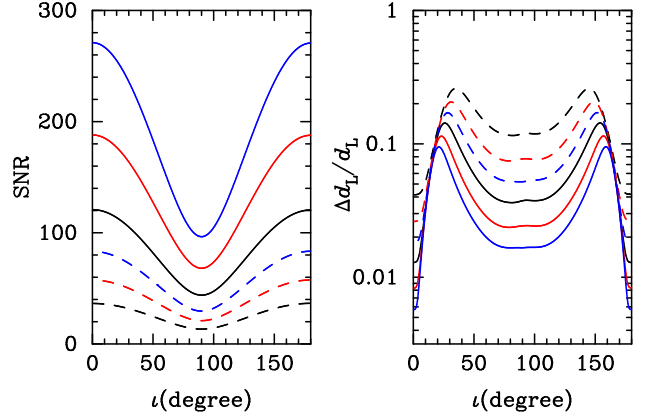}}
\end{center}
\caption{
Dependence of the resulting SNR (left panel) and relative error of luminosity distance (right panel) on $\iota$, obtained for the SMBBH candidate SDSS J164452.71+430752.2 with $M_{\bullet\bullet}=1.413\times10^{10}M_{\odot}$ and $f_0=1.16\times10^{-8}$Hz. In each panel, the blue, red, and black solid lines show the results for the cases with $(N_{\rm p}, q)= (1026, 1)$, $(500, 1)$, and $(200, 1)$, respectively, while blue, red, and black dashed lines show the results for the cases with  $(N_{\rm p}, q)= (1026, 0.1)$, $(500, 0.1)$, and $(200, 0.1)$, respectively.
The assumed ``PTAs'' here monitors $1026$ pulsars with a cadence of $2$\,weeks and timing noise RMS of $100$\,ns.
}
\label{f2}
\end{figure}

\begin{figure}
\setlength{\abovecaptionskip}{0.cm}
\setlength{\belowcaptionskip}{-0.cm}
\begin{center}
\centerline{\includegraphics[scale=0.32,angle=-90]{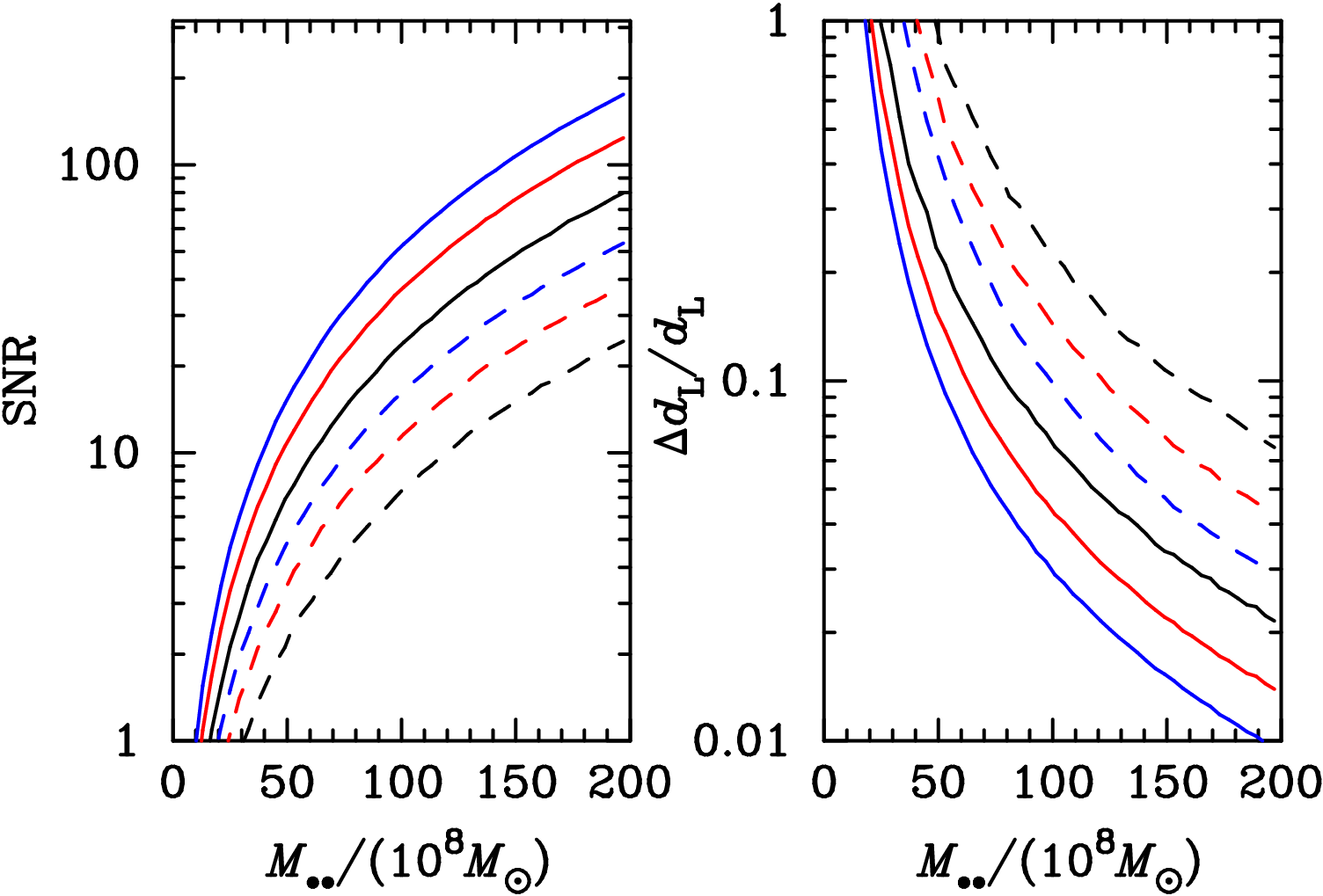}}
\end{center}
\caption{Dependence of the resulting SNR and $\Delta d_{\rm L}/d_{\rm L}$ on the total mass of the system $M_{\bullet\bullet}$ with $f_0=1.157\times10^{-8}$Hz and $\iota=\pi/2$. Legend for the lines are similar to those in Fig.~\ref{f2}. The assumed ``PTAs'' here monitors $1026$ pulsars with a cadence of $2$\,weeks and timing noise RMS of $100$\,ns.}
\label{f3}
\end{figure}

\begin{figure}
\begin{center}
\setlength{\abovecaptionskip}{0.cm}
\setlength{\belowcaptionskip}{-0.cm}
\includegraphics[scale=0.32,angle=-90]{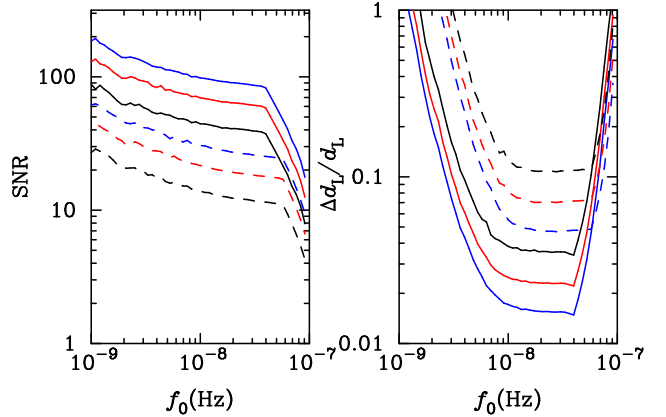}
\end{center}
\caption{ Dependence of the resulting SNR and $\Delta d_{\rm L}/d_{\rm L}$ on the GW frequency of the system with
$M_{\bullet\bullet}=1.413\times10^{10}M_{\odot}$ and $\iota=\pi/2$. Legend for the lines are similar to those in Fig.~\ref{f2}. The assumed ``PTA'' here monitors $1026$ pulsars with a cadence of $2$\,weeks and timing noise RMS of $100$\,ns.}
\label{f4}
\end{figure}

We will first investigate some SMBBH candidates with typical period $\sim 1-10$\,year that are available in the literature, as the main GW sources in the PTAs frequency band are SMBBHs. Taking one of them, SDSS J164452.71+430752.2 at redshift $z=1.715$, as an example, we investigate the effects of inclination angle $\iota$ and mass ratio $q$ on the SNR of its GW signal (see Eq.~\ref{eq:SNR}) and relative error of luminosity distance $\Delta d_{\rm L}/d_{\rm L}$. The total mass of this SMBBH is estimated to be $M_{\bullet\bullet} \sim 1.41\times10^{10}M_{\odot}$ \citep{Shen2008}, and the GW radiation from it is almost monochromatic with a frequency of $f_0 \simeq 1.16\times10^{-8}$Hz if it is on a circular orbit.
Note that in the calculation of $\Delta d_{\rm L}/d_{\rm L}$, we adopt the Fisher matrix analysis, for which $\alpha$ and $\delta$ are fixed as it were accurately determined by its EM counterparts, but not excluded in the analysis if not otherwise stated, and we also further consider the case if the information about the total mass and mass ratio of SMBBH can be given by EM observations. In our calculations, we consider the cases with different numbers of MSPs, i.e. $N_{\rm p}=1026$, $N_{\rm p}=500$ and $N_{\rm p}=200$, respectively. Our main results are plotted in Figures~\ref{f2}, \ref{f3}, and \ref{f4}, respectively.

Figure~\ref{f2} shows the resulting SNR and relative error of luminosity distance $\Delta d_{\rm L}/d_{\rm L}$ as a function of $\iota$. As seen from this Figure, a larger $q$ for a system with given total mass, correspondingly a larger $M_{\rm c}$, results in a larger SNR and a smaller $\Delta d_{\rm L}/d_{\rm L}$. If $\iota=0$, i.e.,  the GW source is face-on, the resulting SNR is then the largest and the GW signal can be easier detected, the resulting $\Delta d_{\rm L}/d_{\rm L}$ show that the luminosity distance can be well determined for face-on or edge-on source and this value will be peak at angle in a range of  $\iota \in (20^{\circ}, 50^{\circ})$ or $(130^{\circ}, 160^{\circ})$. These results are consistent with the results for the ground-based GW detectors \citep[e.g.,][]{2019PhRvX...9c1040A}.

Figures~\ref{f3} and \ref{f4} show the dependence of the resulting SNR and $\Delta d_{\rm L}/d_{\rm L}$ on the total mass and GW frequency of the system with fixed $\iota=\pi/2$, respectively. As seen from these Figures, the larger the total mass of the system, the larger the resulting SNR and the smaller the resulting $\Delta d_{\rm L}/d_{\rm L}$;  the larger the initial GW frequency $f_0$, the smaller the resulting SNR and the smaller the resulting $\Delta d_{\rm L}/d_{\rm L}$, except at $f_0 \gtrsim 4\times 10^{-8}$\,Hz. The larger $f_0$ means the smaller semimajor axis of the SMBBH system and the lager change rate of the frequency, which leads to a better determination of the luminosity distance $d_{\rm L}$, but a decrease of SNR as it $\propto f^{-1/3}$. The rapid decrease of SNR at $f_0 \gtrsim 4\times 10^{-8}$\,Hz) is due to that those SMBBH systems have a merger timescale $\tau_{\rm GW}$ less than the observation period (e.g., $T_{\rm obs} = 10$\,yr) and thus  SNR$\propto \tau_{\rm GW}^{1/2} \propto f^{-\gamma}$ with $\gamma > 4/3$ as SMBBHs at this stage are not continuous GW sources and $f$ increases fast.

\begin{figure}
\begin{center}
\setlength{\abovecaptionskip}{0.cm}
\setlength{\belowcaptionskip}{-0.cm}
\includegraphics[scale=0.50, angle=-90]{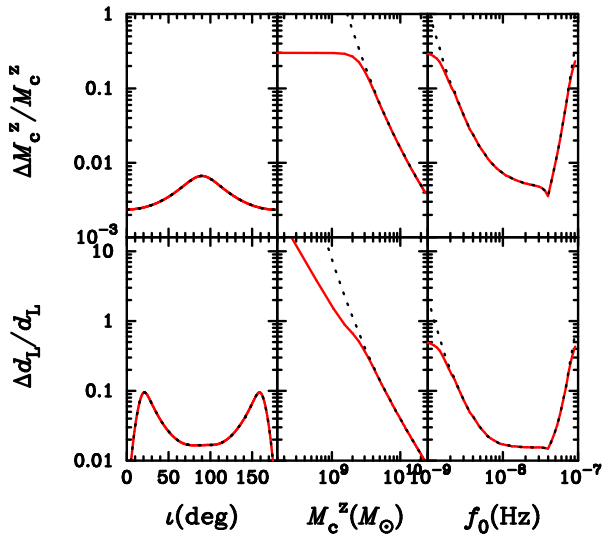}
\end{center}
\caption{
Dependence of the resulting $\Delta M^z_{\rm c}/M^z_{\rm c}$ and $\Delta d_{\rm L}/d_{\rm L}$ on inclination angel (left panel), redshifted chirp mass $M_{\rm c}^z$ (middle panel) and frequency $f_0$ (right panel) expected from the ``PTAs observations'' of an sets of SMBBH systems with redshift $z=1.715$ and mass ratio $q=1$. In each panel, black dotted line shows the result obtained by considering all the eight parameters of the GW source as free ones in the Fisher Matrix, while the red solid line shows the case by adding additional information on the redshifted chirp mass, following a Gaussian distribution with a scatter of $\sigma \ln M_{\rm c}^{\rm z}=0.3$ in the Fisher Matrix. The assumed ``PTAs'' here monitors $1026$ pulsars with a cadence of $2$\,weeks and timing noise RMS of $100$\,ns.
}
\label{f5}
\end{figure}

Additional information on the (total) mass and mass ratio of the PTAs SMBBHs may be obtained from the EM measurements, and the degeneracy between chirp mass and frequency can thus be broken, leading to a significant improvement of the $d_{\rm L}$ estimation.
Figure~\ref{f5} shows the errors for luminosity distance estimates $\Delta d_{\rm L}$
from the PTA data only (black dotted lines in each panel) and those for the estimates from the PTA data with additional information on the redshifted chirp mass $M_{\rm c}^z$ (red solid line in each panel), respectively. In each panel, the black dotted lines show the results obtained by considering all the eight parameters of the GW source as free ones in the Fisher Matrix, while the red solid line show the cases by adding additional information on the redshifted chirp mass, following a Gaussian distribution with a scatter of $\sigma \ln M_{\rm c}^{\rm z}=0.3$ in the Fisher Matrix. Left panel shows $\Delta d_{\rm L}/ d_{\rm L}$ against the inclination angle $\iota$ of the system. Middle panel shows that against the input total mass of the PTAs SMBBHs, and right panel shows $\Delta d_{\rm L}/ d_{\rm L}$ against the input $f_0$ of the PTAs SMBBHs.
According to this Figure, if the errors of the redshifted chirp mass can be obtained from the EM observations, e.g., via the reverberation mapping method (independent of the cosmological model), with high precision, the measurement errors in $d_{\rm L}$ can be significantly suppressed, especially when $f_0  \lesssim 10^{-8}$\,Hz and $M_{\bullet\bullet}$ in the range of $\sim 10^{9}-10^{10}M_{\odot}$.
Since the number density of SMBBHs with $M_{\bullet\bullet} \sim 10^9M_\odot$ is much larger than that with $M_{\bullet\bullet} \sim 10^{10}M_\odot$, the number of PTAs SMBBHs that can be used as standard sirens to constrain cosmology may increase significantly if additional information can be provided by electromagnetic observations and thus lead to better constraints on cosmology.

\section{PTAs SMBBHs as standard sirens}
\label{sec:mbbhs}

It is longly anticipated to detect the GWs from SMBBHs via PTAs, but not any GW from a single SMBBH was detected in the past. Electromagnetic observations do suggest a number of SMBBH candidates but it is not sure how many of those SMBBH candidates are true SMBBHs. In this Section, we will first consider those SMBBH candidates and then consider a mock sample of SMBBHs obtained from a simple galaxy merger model, in order to investigate possible constraints on the dark energy that may be obtained if assuming all those SMBBHs are true SMBBHs via future PTAs observations.

One of the crucial points for using PTAs SMBBHs to constrain cosmological parameters is to detect these systems by electromagnetic waves and get  redshift measurements. Since SMBBHs are formed by mergers of galaxies, nuclear activities are believed to be triggered in many of such SMBBH systems with distinct signatures. For this reason, many SMBBH systems should be detectable via electromagnetic wave, which offers redshift measurement. Indeed, there have been a lot of efforts made in the past several decades to search for SMBBHs through their electromagnetic signatures, including the periodic variation in light curves \citep[e.g.,][]{Graham15, Charisi16, Valtonen08}, double-peaked or asymmetric broad emission lines \citep[e.g.,][]{Liu2014, Li16, 2019MNRAS.482.3288G, Li19}, UV-optical deficit in the spectral energy distribution \citep{Yan15, Zheng16}, etc. These efforts have been resulted in more than one hundred SMBBH candidates, though more efforts are still needed to confirm them. In the following calculations, therefore, we assume that the redshift of all PTAs SMBBHs can be obtained by electromagnetic observations.\footnote{Note that some SMBBHs may be quiescent, which are indeed not easy to be detected electromagnetically. Ignoring this should not affect our conclusion qualitatively.}
Note that we also ignore the effects due to dynamical environments of active SMBBHs on the orbital decay in addition to the GWs. This should also lead to some uncertainties in the use of PTAs SMBBHs as standard sirens, but may be corrected by detailed studies of each individual source.

\subsection{SMBBH candidates from EM observations}
\label{subsec:mBBHcand}

We adopt the current available sample for SMBBH candidates ($154$) obtained from various characteristic signatures with estimates on total masses $M_{\bullet\bullet}$ and orbit period $T$. Among these SMBBH candidates, most ($149$) are obtained via periodic variations in their light curves
\citep{Graham15, Graham15b, Charisi16}, others are Mrk\,231 from \citet{Yan15}, NGC\,5548 from \citet{Li16}, OJ\,287 from \citet{Valtonen08}, SDSS J0159+0105 from \citet{Zheng16}, and Ark 120 from \citet{Li19}.
The mass ratio $q$, inclination angle $\iota$, polarization angle $\psi$ and initial phase $\phi_0$ of most SMBBH candidates are not known, yet. For this reason, we shall consider (1) two  different mass ratio, i.e., $q=1$ and $0.1$, respectively, for those SMBBH candidates with no information on the mass ratio; (2) different $\iota$ values. However, we fix $\psi=0$ and $\phi_0=0$ as these two angles  have no significant effects on the results.
We also assume that those SMBBH candidates are all on circular orbits. Therefore, its orbit frequency $f_{\rm orb}$ is $\frac{1}{2\pi}\sqrt{G M_{\bullet\bullet}/a^3}$, where $a$ is the orbital radius of the system, the GW frequency is $f_{\rm GW}=2f_{\rm orb}$.\footnote{In principle, eccentric SMBBHs can be considered, though it is more complicated than circular ones as the GW emission is not monochromatic and additional assumption needs to be made for the eccentricity distribution. We assume all SMBBHs are on circular orbits for simplicity.}
The green stars in Figure~\ref{fig:f1} mark the position of those SMBBH candidates.

\begin{figure*}
\begin{center}
\setlength{\abovecaptionskip}{0.cm}
\setlength{\belowcaptionskip}{-0.cm}
\centerline{\includegraphics[scale=0.64,angle=-90]{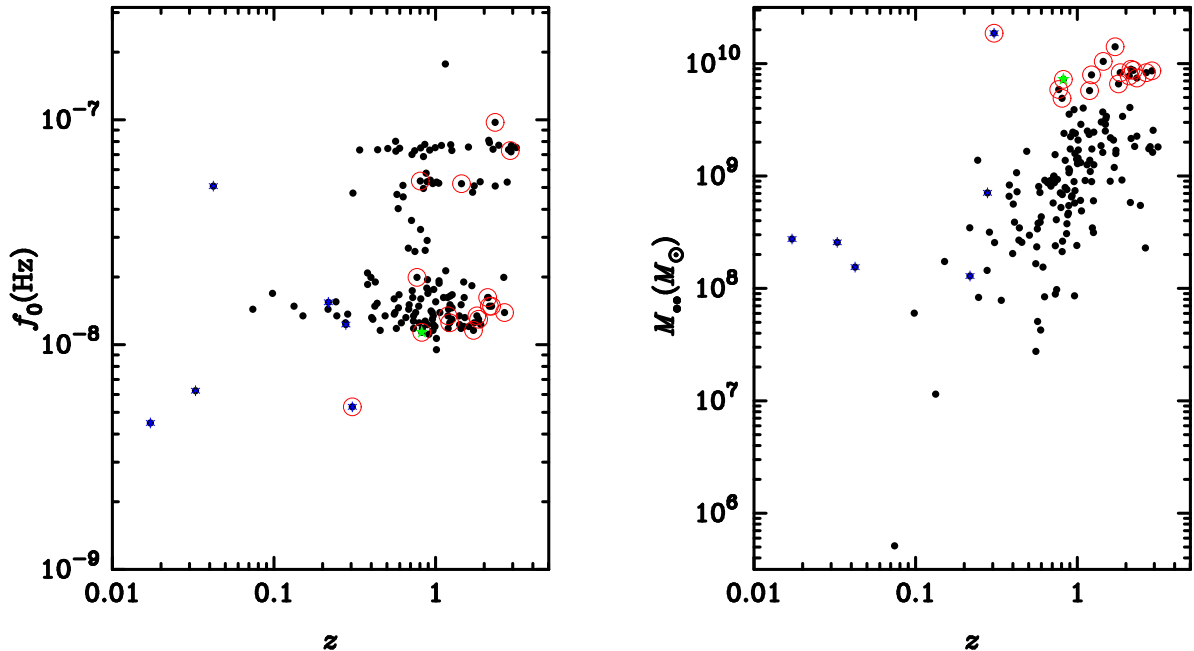}}
\end{center}
\caption{Distribution of SMBBH candidates (black points) in the  $f_0-z$ plane (left) and in the $M_{\bullet\bullet}-z$ plane (right). In each panel, the red circles mark those SMBBH candidates that their expected SNR $\rho > 10$ and $\Delta d_{\rm L} /d_{\rm L} <1.0$ if  monitored by a PTAs with $N_{\rm p}=1026$, a cadence of $2$\,weeks, and timing noise RMS $\sigma_t=100$\,ns. The SMBBH candidates are assumed to be equal mass ($q=1$) and have $\iota=\pi/2$.}
\label{f6}
\end{figure*}

\begin{figure*}
\begin{center}
\setlength{\abovecaptionskip}{0.cm}
\setlength{\belowcaptionskip}{-0.cm}
\centerline{\includegraphics[scale=0.64,angle=-90]{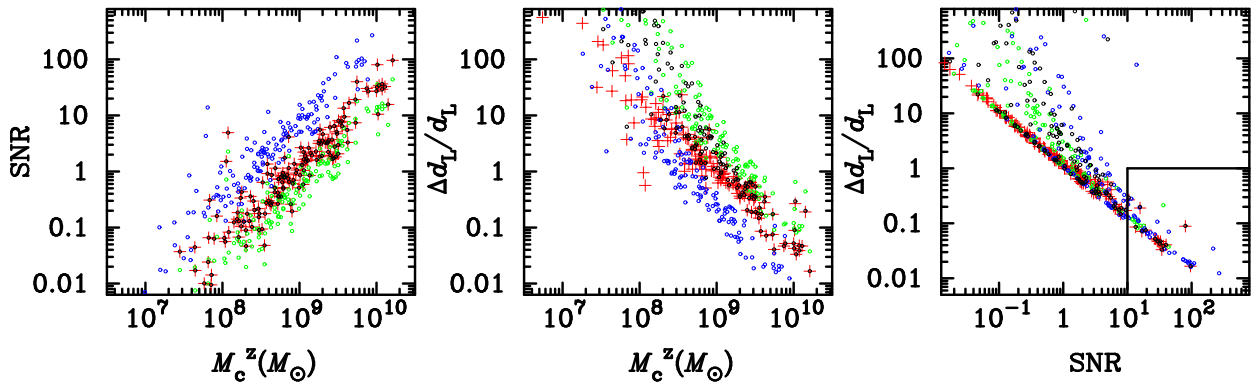}}
\end{center}
\caption{Expected signal-to-noise ratio (SNR) and luminosity distance measurement errors of those SMBBH candidates by a PTAs with settings of $(N_{\rm p}$, $\iota$, $\sigma \ln M_{\rm c}^z$ )= $(1026, \pi/2, \cdots)$ (black circles) , $(1026, \pi/2, 0.3)$ (red $+$), $(200, \pi/2, \cdots)$ (green circles), and $(1026, \pi/10, \cdots)$ (blue circles), respectively. Note that the results shown here are obtained by assuming $q=1$ for all SMBBH candidates.}
\label{f7}
\end{figure*}

We consider a number of PTAs settings on the number of usable MSPs ($N_{\rm p}$), pulsar timing noise RMS $\sigma_{\rm t}$, and the cadence $\Delta t$, which may be possible in the SKA era (as listed in Tables~\ref{table1} and \ref{table2}). With these PTAs settings, the expected SNR for each SMBBH candidate can be calculated according to Equation~(\ref{eq:SNR}).
Figure~\ref{f6} shows these SMBBH candidates in the $M_{\rm tot}-z$ plane and in the $f_0-z$ plane, in which the objects with SNR$>10$ are marked with red circles. It is evident that only the SMBBH candidates with large $M_{\bullet\bullet}$ (or correspondingly large chirp mass $M_{\rm c}$) are detectable in the SKA-PTA era (e.g., with SNR\,$>10$). However, the dependences on frequency $f_0$ and redshift $z$ are not significant. Figure~\ref{f7} shows the expected SNR and the precision of luminosity distance measurements ($\Delta d_{\rm L}/d_{\rm L}$) of those SMBBH candidates (assuming $q=1$) from PTAs with different settings. The determination of $d_{\rm L}$ are quite good for sources with large $M_{\bullet\bullet}$ and high SNR (e.g., $>10$). For these sources, if a mass ratio of $q=1$ is assumed, then there will be more sources that can have relatively accurate $d_{\rm L}$ measurements (with small $\Delta d_{\rm L} /d_{\rm L}$), however, the number of such sources is substantially smaller if a smaller mass ratio is adopted ($q=0.1$; see Table~\ref{table1}) because of much weaker GW signals.

We only choose those sources with SNR $\rho>10$ and $\Delta d_{\rm L} /d_{\rm L}<1$ to estimate the precision of constraint on the dark energy EoS $\Delta w$ that may be obtained from PTAs observations. The reason is that only sources with sufficiently high SNR can be detected by PTAs and only sources with sufficiently small $\Delta d_{\rm L} /d_{\rm L}$ are useful to obtain strong constraint on dark energy.
If the sampling rate can be enhanced (smaller cadence $\Delta t$; see Eq.~\ref{eq:SNR}, SNR approximately $\propto N^{1/2} \propto 1/\sqrt{\Delta t}$) or the timing noise RMS can be suppressed (smaller $\sigma_{\rm t}$; see Eq.~\ref{eq:SNR}, SNR $\propto 1/\sigma_{\rm t}$), the SNR and luminosity distance measurements can both be improved and thus a better constraint on the dark energy EoS can be obtained. Therefore, according to the results obtained above and those listed in Table~\ref{table1}, we conclude that only those SMBBHs with large chirp mass could be treated as standard sirens and can be used to get a strong constraint on the EoS of dark energy $\Delta w \sim 0.04-0.06$ under the most optimistic conditions (see the last row in Table~\ref{table1}, which are estimated by using Eq.~(\ref{eq:dw}) as described in Section~\ref{sec:method}).

\begin{table}
%
%
\caption{Expected constraints on the Equation of State of dark energy from current available SMBBH candidates.}
\centering
\begin{tabular}{cccc  c c cc c  }   \hline
 $\Delta t$ & $\sigma_t$ & \multirow{2}{*}{$N_{\rm P}$} & \multirow{2}{*}{$\sigma {\rm ln}M_{\rm c}^z$} & \multicolumn{2}{c}{$q=1$} & &
 \multicolumn{2}{c}{$q=0.1$} \\ \cline{5-6} \cline{8-9}
 (week) & (ns) & & & $N_{\rm s}$ & $\Delta w$ & &  $N_{\rm s}$ & $\Delta w$ \\
 \hline
2 & 100 & 200  &-&   11 & 0.17 & & 2 & 0.63  \\ \hline
2 & 100 & 500  & - & 14 & 0.12 & & 2 & 0.44  \\ \hline
2 & 100 & 1026 & - & 16 & 0.096 & & 5 & 0.26  \\ \hline
2 & 100 & 1026 &0.5& 16 & 0.096 & & 5 & 0.26 \\ \hline
2 & 100 & 1026 &0.3& 16 & 0.096 & & 5 & 0.26 \\ \hline
1 & 100 & 1026 &0.3& 21 & 0.079 & & 12 & 0.16  \\ \hline
2 & 50  & 1026 &0.3& 27 & 0.065 & & 15 &  0.12 \\ \hline
1 & 50  & 1026 &0.3& 34 & 0.055 & & 16 & 0.097 \\ \hline
1 & 20  & 1026 &0.3& 65 & 0.036 & & 30 & 0.056 \\ \hline
\end{tabular}
\tablecomments{The numbers listed in the sixth and eighth columns represent the uncertainty of the constraint on the dark energy Equation of State.
Results listed in the fifth and sixth (or seventh and eighth) columns are obtained by assuming that all the SMBBHs candidates have a mass ratio of $q=1$ (or $q=01.$). $N_{\rm s}$ is the number of the sources which satisfy the condition of SNR$>10$ and $\Delta d_{\rm L}/d_{\rm L}<1.0$. In the calculations, all SMBBH candidates are also assumed to have $\iota=\pi/2$. Note that in the first three cases no information from electromagnetic observations on the mass is added, so the forth column is shown as a hyphen.}
\label{table1}
\end{table}


\subsection{Mock SMBBHs from a simple galaxy merger model}
\label{subsec:mock}

In this section, we generate a mock sample of SMBBHs according to a simple galaxy merger model since SMBBHs were generally formed via galaxy mergers. The main input quantities for this model are the cosmic merger rate of galaxies and the relationship between MBH mass and host galaxy properties, which can all be given by observations. We assume that the time delay between the merger of two galaxies and formation of a central SMBBHs is short ($\sim 1$\,Gyr) comparing with the cosmic time \citep[e.g., for detailed dynamical merging processes see, e.g.,][]{Yu02, Chen2019}, therefore it can  be ignored. After the SMBBH enters into the PTA band, its orbital decay dominates by the GW radiation and thus the environmental effect can be also ignored. We further assume that SMBBHs are well circularized at frequencies $\gtrsim 10^{-9}$\,Hz and thus considerations of the eccentricities of SMBBHs are not needed in the following calculations. For the central SMBBHs, the masses of its two components can be estimated according to the relationship between MBH mass and galaxy properties \citep{kormendy13}, such as stellar mass.

The comoving number density of SMMBHs is controlled by the galaxy-galaxy merger rate density, the relationship between MBH mass and galaxy properties, and the residential time of SMBBHs at different semimajor axis (and thus different GW frequencies under the GW decay).
For SMBBHs at redshift $z$ with total mass $M_{\bullet\bullet}$ and mass ratio $q$, its comoving number density per unit comoving volume $V$, $q$, $\log M_{\bullet\bullet}$, and logarithmic GW observed frequency $\log f$ can be estimated as
\bea
\frac{dN(z,M_{\bullet\bullet},q,f)}{dV d\log f dq d \log M_{\bullet\bullet}}&=&
\int d\mu_{\rm gal} \int_0^{\infty} P(M_{\bullet\bullet}|\bar{M}_{\bullet\bullet}(M_{\rm gal},z))\nonumber\\
&&\times\frac{dN_{\rm mrg}}{d\mu_{\rm gal}dt}(M_{\rm gal},\mu_{\rm gal},z) \delta (q-\mu_{\rm gal}) \nonumber\\
&&\times\Phi(M_{\rm gal}, z) \frac{\Delta t}{\Delta \log f}dM_{\rm gal}.
\label{eq:dNdV}
\eea
Here $P(M_{\bullet\bullet}|\bar{M}_{\bullet\bullet}(M_{\rm gal},z))$ is assumed to be a Gaussian probability distribution function with a scatter of $\Delta M_{\bullet}$ that describes the distribution of the true MBH mass around the mean value $\bar{M}_{\bullet\bullet}(M_{\rm gal},z)$ obtained from the $M_{\bullet\bullet} - M_{{\rm bulge}}$ relationship given in \citet{kormendy13}, i.e., $\bar{M}_{\bullet\bullet} =0.49 \times 10^9(M_{\rm gal} /10^{11} M_{\odot})^{1.17} M_{\odot}$ with a scatter of $\Delta {\rm log} M_{\bullet\bullet}=0.3$. The galaxy-galaxy merger rate $\frac{dN_{\rm mrg}}{d\mu_{\rm gal}dt}(M_{\rm gal},\mu_{\rm gal},z)$ is obtained from the merger trees for galaxies in the Illustris Simulation by directly tracking the baryonic content of subhalos \citep[][]{Rodriguez15},
where $M_{\rm gal}$ is the mass of the merger remnant galaxy, $\mu_{\rm gal}$ the mass ratio of the two progenitor galaxies. $\Phi(M_{\rm gal}, z)$ is the galaxy stellar mass function, for which we adopt the estimates from \citet[][see Table\,4 therein]{lopes17}. Note that for simplicity we assume $q = \mu_{\rm gal}$ and ignore the growth of MBHs before the formation of SMBBHs so that the total mass of the progenitor SMBBH equals the mass of MBH in the merger remnant galaxy. Since the orbital decay is governed by the GW radiation, the time period $\Delta t$ that an SMBBH stays at the frequency band from $f$ to $f+\Delta f$ is given by
\be
\Delta t = \frac{8}{3}\frac{{\rm ln}10(\pi f(1+z))^{-8/3}}{\frac{256}{5}(GM_{\rm c}/c^3)^{5/3}} \Delta \log f.
\ee

We randomly generate mock SMBBHs in the parameter space of $z \in (0, 4)$, $f_0 \in (10^{-9}$\,Hz, $10^{-7}$\,Hz),  $M_{\bullet\bullet} \in (10^7\msun, 10^{11}\msun)$, and $q \in (0.01, 1)$, according to Equation~(\ref{eq:dNdV}) by integrating it over the cosmic volume. For each mock SMBBH, its sky location is randomly set by assuming an uniform distribution of SMBBHs on the two dimensional sky. Its orbital inclination relative to the line of sight is randomly set over the range of $\cos\iota \in [-1, 1]$. We also fix the polarization angle $\psi = 0$ and the initial phase angle $\phi_0 = 0$, for simplicity. With all the parameters set above, i.e., $(z, \alpha, \delta, \iota, M_{\bullet\bullet}, q, f_0, \psi, \phi_0)$, for each mock SMBBH, the SNR of its GW signal can be obtained for any given PTAs. In the calculations of SNRs, we assume that the sky position of those SMBBHs can be obtained by the EM signals from those SMBBHs, similar to the SMBBH candidates studied in Section~\ref{subsec:mBBHcand}.

\begin{figure*}
\begin{center}
\setlength{\abovecaptionskip}{0.cm}
\setlength{\belowcaptionskip}{-0.cm}
\centerline{\includegraphics[scale=0.60, angle=-90]{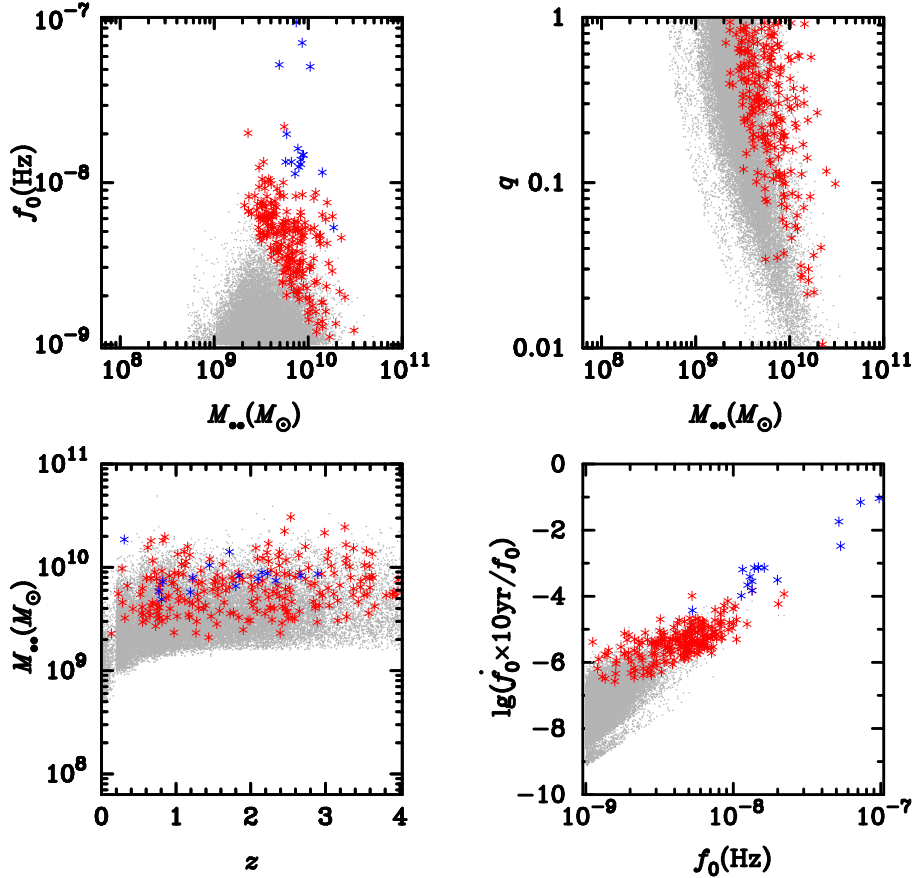}}
\end{center}
\caption{Distribution of SMBBHs in the $f_0$ vs $M_{\bullet\bullet}$ plane (top-left), $q$ vs $M_{\bullet\bullet}$ plane (top-right),  $M_{\bullet\bullet}$ vs $z$ plane (bottom-left), and   $\dot{f_0}/f_0$ vs $f_0$ plane (bottom-right). In each panel, the gray dots show all mock SMBBHs in the simulated sample with SNR $\rho>10$, red stars show the mock SMBBHs with $\Delta d_{\rm L} / d_{\rm L} < 1$, while blue stars  (except in the top-right panel) indicate those SMBBH candidates (assuming $q=1$) with $\rho>10$ and $\Delta d_{\rm L} / d_{\rm L} <1$. The SNR for each mock SMBBH is estimated by assuming a PTAs observations with $N_{\rm p}=1026$, cadence $\Delta t=2$\,weeks and $\sigma_t=100$\,ns . }
\label{f8}
\end{figure*}

\begin{figure*}
\begin{center}
\setlength{\abovecaptionskip}{0.cm}
\setlength{\belowcaptionskip}{-0.cm}
\centerline{\includegraphics[scale=0.64,angle=-90]{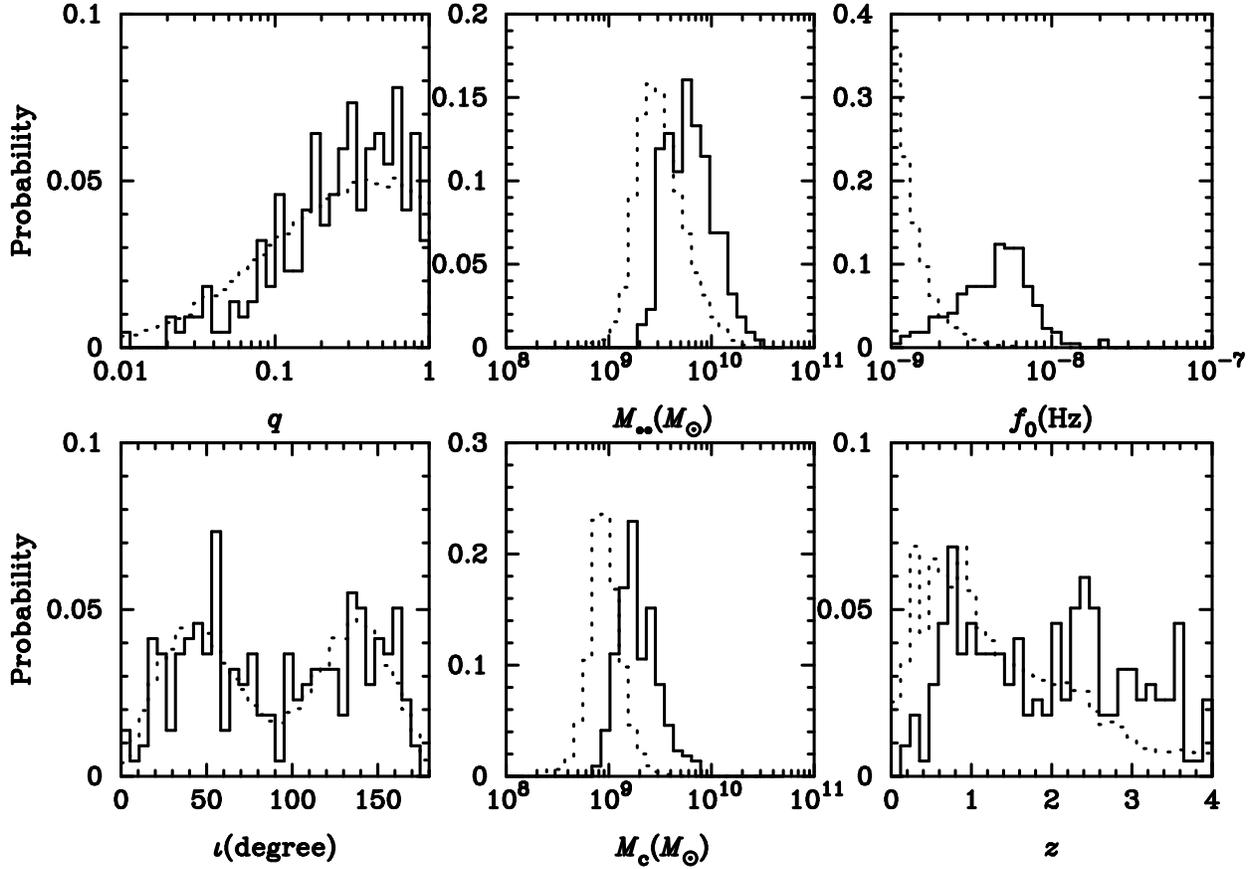}}
\end{center}
\caption{Distributions of simulated SMBBHs against  mass ratio $q$ (top left), total mass $M_{\bullet\bullet}$ (top middle), initial frequency $f_0$ (top right), polarization angle $\iota$ (bottom left), chirp mass $M_{\rm c}$ (bottom middle), and redshift (bottom right). In each panel, the dotted histogram shows the distribution of all mock SMBBHs in the sample (with $\rho>10$), while the solid histogram shows that for those mock SMBBHs with SNR $\rho>10$ and $\Delta d_{\rm L}/d_{\rm L}<1$.
}
\label{f9}
\end{figure*}

\begin{figure*}
\begin{center}
\setlength{\abovecaptionskip}{0.cm}
\setlength{\belowcaptionskip}{-0.cm}
\centerline{\includegraphics[scale=0.75, angle=-90]{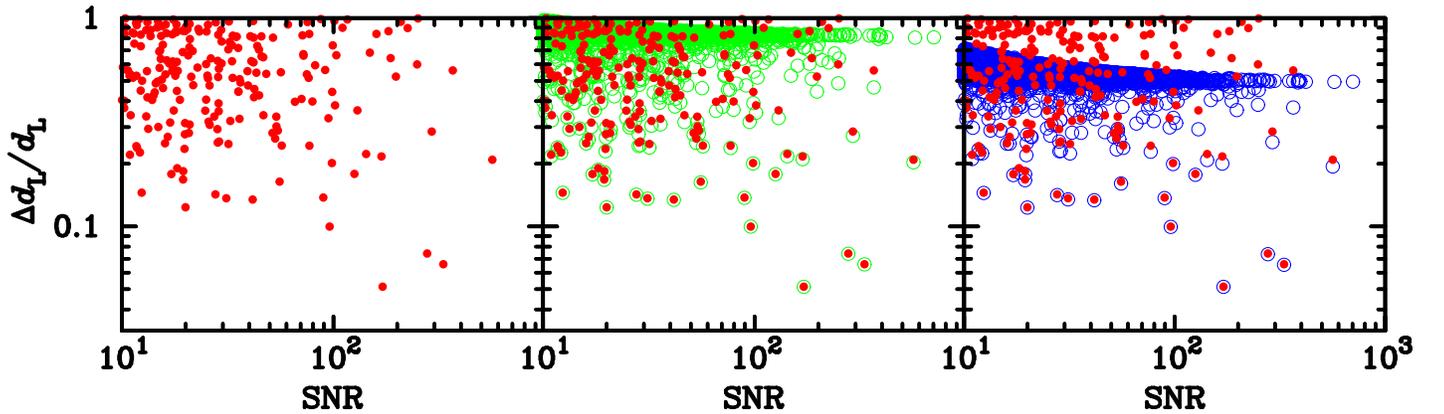}}
\end{center}
\caption{Expected errors of luminosity distance measurements for those mock SMBBH (with $\rho>10$) by a PTAs with pulsar number $N_{\rm p}=1026$, cadence $\Delta t=2$\,weeks, and timing RMS noise $\sigma_t=100$\,ns. Left, middle, and right panels show the results obtained by assuming the chirp mass as a free parameter (red symbols), a free parameter with $1-\sigma$ error of $\sigma \ln M_{\rm c}^z = 0.5$ (green circles), and $0.3$ (blue circles), respectively. This Figure shows that the measurements of luminosity distances from GW signals of many SMBBHs can be significantly improved if errors on chirp mass of SMBBH systems can be obtained from electromagnetic observations and thus the number of SMBBHs that can be used as standard sirens significantly increases with increasing precision of such additional information.}
\label{f10}
\end{figure*}

We also consider a number of possible PTAs settings in the SKA era as that in the previous Section~\ref{subsec:mBBHcand} (see Tables~\ref{table1} and \ref{table2}). We adopt two different SNR thresholds for those mock SMBBHs to be adopted in obtaining constraint on the dark energy EOS, i.e., $\rho>10$ and $>50$, respectively. For the selected samples, using the Fisher matrix analysis, we derive the value of $\Delta  d_{\rm L}/ d_{\rm L}$ and then obtain the constraint on the EoS of dark energy in different conditions as show in Table \ref{table2}. We illustrate our main results in Figures~\ref{f8}, \ref{f9}, and \ref{f10} as follows.

Figure~\ref{f8} shows the distribution of  both mock SMBBHs and observational SMBBH candidates (with $\rho>10$) on the plane of $f_0-M_{\bullet\bullet}$ (top left panel), $q-M_{\bullet\bullet}$ (top right panel), $M_{\bullet\bullet}-z$ (bottom left panel), and $\log (\dot{f}_0\times 10{\rm yr}/f_0)-f_0$ (bottom right panel), respectively. The PTAs adopted here has the settings of $N_{\rm p}=1026$, $\Delta t=2$\,weeks, and $\sigma_t=100$\,ns, as shown by the third row in Table~\ref{table2}. Gray dots shown in this Figure represent all the mock SMBBHs with SNR $\rho>10$, while the red and blue stars mark the mock SMBBHs and observational SMBBH candidates with $\rho>10$ and $\Delta d_{\rm L}/ d_{\rm L} <1$. Only a small fraction of SMBBHs with $\rho>10$ that can have relatively good distance measurements $\Delta d_{\rm L}/ d_{\rm L}< 1$, which can be used as the standard sirens.\footnote{The four blue stars are quite faraway from the mock SMBBHs (top-left panel and bottom-right panel in Figure~\ref{f8}), which might be due to (1) some of the SMBBH candidates are not real SMBBHs, (2) selection effect, and (3) our model under predict the abundance of SMBBHs at high frequency, though it is not likely.} The SMBBHs in this sample have relatively large $M_{\bullet\bullet}$ ($\gtrsim 3\times 10^9 M_\odot$, bottom left panel), relatively large $f_0$ (top left panel), and relatively large $\dot{f_0} / f_0$ (bottom right panel).

For comparison, Figure~\ref{f9} shows the distributions of the total mass $M_{\bullet\bullet}$, mass ratio ($q$), chirp mass $M_{\rm c}$, frequency $f_0$, inclination angle ($\iota$), and redshift ($z$) of SMBBHs in the selected SMBBH sample (red stars in Fig.~\ref{f8}) and the whole sample (with $\rho>10$, grey points in Fig.~\ref{f8}). It is clear that the distributions of $M_{\bullet\bullet}$, $f_0$, $M_{\rm c}$ in the selected sample are biased to the high-value ends from those in the parent SMBBH sample. The redshift distribution of SMBBHs in this selected sample is also different from that in the parent sample. The flat redshift distribution of SMBBHs in the selected sample  suggests that usable PTAs SMBBH standard sirens can be detected at high redshift and thus offers a good tool to probe high redshift universe. The distribution of $\iota$ in the selected sample is more or less the same as that in the parent sample. The expected SNRs for SMBBHs with $\iota \sim 90^\circ$ are relatively smaller, therefore, there lacks of SMBBHs with such $\iota$ in the selected sample (red stars). The lacks of SMBBHs with $\iota\sim 0^\circ$ or $\sim 180^\circ$ are mainly caused by the randomly distribution of orbital orientation. The fraction of sources in frequency range $10^{-8}-10^{-7}$\,Hz is extremely small (top-right panel of Fig.~\ref{f9}) mainly because the higher frequency corresponds to the smaller SMBBH separation ($a$) and thus much smaller residential time (as it proportional to $a^4$). Most of the selected SMBBHs have $f_0 <10^{-8}$\,Hz, where the frequency change rate is too small to be well measured by the PTAs and thus hinder the distance measurement with high precision.

If the SKA-PTA is conservatively set as $\delta t=2$\,weeks and $\sigma_t=100$\,ns, Using the mock SMBBHs with $\rho>10$ (or $>50$) and $\Delta d_{\rm L}/ d_{\rm L} <1 $ (red stars in Figure~\ref{f8}, only $211$ with $\rho>10$ and $65$ with $\rho >50$) as standard sirens, we can get an estimate of its constraining power on the dark energy for future PTAs (see the third row in Table~\ref{table2}). Apparently, the obtained constraint on the EoS of dark energy is about $\Delta w \sim 0.1-0.2$, not very tight. If the SKA-PTA is optimistically set as  $\delta t=1$\,weeks and $\sigma_t=20$\,ns, the mock SMBBHs with $\rho>10$ and $>50$ can be upto to $3000$ and $1500$, respectively, with which the constraint on the EoS of dark energy can be achieved to $\sim 2\%$ level. For comparison, the potential constraints of EoS of dark energy by SN Ia or by weak lensing observations are expected to be around $\Delta w \sim 0.01$ \citep{Albrecht2006}.

Electromagnetic observations of many PTAs SMBBHs may enable the measurements of their physical properties, which may be combined with the PTAs GW signal to improve the measurements of $d_{\rm L}$. Such information include the total mass and mass ratio of an SMBBH system, which can be obtained by the reverberation mapping technique and detailed analysis of its spectral energy distribution. Therefore, an estimation on the error of the chirp mass is possible, which is not cosmological model dependent. Here we assume that the chirp mass accuracy can be obtained from electromagnetic observations to the order of $\sigma \ln M_{\rm c}^z \sim 0.5$ or $0.3$ and consider the improvement of the constraints on the EoS of dark energy from the PTAs SMBBHs. Figure~\ref{f10} shows that the number of mock SMBBHs with $\rho >10$ and $\Delta d_{\rm L} /d_{\rm L}<1$ increases significantly if put additional information on the chirp mass into the analysis (see Tables~\ref{table1} and \ref{table2}). For those sources with $\Delta d_{\rm L}/d_{\rm L} \lesssim 0.2-0.3$, the improvements in $\Delta d_{\rm L} /d_{\rm L}$ are negligible as further improvements require much more accurate mass measurements than the electromagnetic observations can give (see the middle and right panels of Fig.~\ref{f10}). As seen from the last few rows in Tables~\ref{table1} and \ref{table2}, the constraints on the EoS of dark energy ($\Delta w$) can reach to $\lesssim 1\%$ level in the most optimistic cases (see last three rows in Table~\ref{table2}) if considering that the error of chirp mass estimates from electromagnetic observation can of SMBBHs can be as accurate as $\sigma \ln M^z_{\rm c} \sim 0.3-0.5$. Such a constraint is quite accurate even comparing with the next generation of large scale structure surveys \citep{Albrecht2006}.

%
\begin{table}
\caption{Expected PTAs constraints on the Equation of State of dark energy from mock SMBBHs.}
\centering
\begin{tabular}{ccccc c cc c }   \hline
 $\Delta t$ & $\sigma_t$\ &
 \multirow{2}{*}{$N_{\rm P}$} & \multirow{2}{*}{$\sigma {\rm ln}M_{\rm c}^z$} & \multicolumn{2}{c}{$\rho>10$} & & \multicolumn{2}{c}{$\rho>50$} \\ \cline{5-6} \cline{8-9}
 (week) & (ns) & &  & $N_{\rm s}$ & $\Delta w$ & & $N_{\rm s}$ & $\Delta w$  \\
 \hline
2 & 100 & 200  & - &  45 & 0.32   & & 9  & 0.45 \\ \hline
2 & 100 & 500  & - &  121 & 0.18   & & 28 & 0.26 \\ \hline
2 & 100 & 1026 & - & 211 & 0.13   & & 65 & 0.18 \\ \hline
1 & 100 & 1026 & - & 362 & 0.090   & & 113 & 0.13 \\ \hline
2 & 50 & 1026 & - & 606 & 0.063   & & 202 & 0.093 \\ \hline
1 & 50 & 1026 & - & 1020 & 0.045   & & 394 & 0.063 \\ \hline
1 & 20 & 1026 & - & 3102 &  0.020  & & 1578 & 0.025 \\ \hline
2 & 100 & 1026 &0.5&33848& 0.026  & &957 &0.12\\ \hline
2 & 100 & 1026 &0.3&33848& 0.017  & &957 &0.085\\ \hline
1 & 100 & 1026 &0.3&66707& 0.0089  & &2183&0.043\\ \hline
2 & 50  & 1026 &0.3&128574& 0.0048  & &4891&0.021\\ \hline
1 & 50  & 1026 &0.3&242352& 0.0026 & &10356&0.011\\ \hline
\end{tabular}
%
%
\label{table2}
\tablecomments{Legend similar to that for Table~\ref{table1}. Results listed in the fifth and sixth (or seventh and eighth) columns are obtained by adopting those mock SMBBHs with $\rho>10$ (or $\rho> 50$) and $\Delta d_{\rm L}/d_{\rm L}<1$. Note that in the first seven cases no information from electromagnetic observations on the mass is added, so the forth column is shown as a hyphen.}
\end{table}

\section{Conclusions and Discussions}
\label{sec:conclusion}ƒ

Nano-hertz frequency GWs from individual SMBBHs are expected to be detected by PTAs in the near future. These ``PTAs detected'' SMBBHs may also be used as standard sirens to probe cosmology. In this paper, we investigate whether such ``PTAs detected'' SMBBHs can be used to obtain independent distance measurements and put strong constraint on the EoS of dark energy. To do this, we adopt the Fisher information matrix for parameter estimations to the expected GW signals from those current available SMBBH candidates and a mock sample of SMBBHs produced from a simple galaxy merger model. We find that the luminosity distance measurements from the GW signals of some SMBBHs with high SNR can be relatively accurate ($\Delta d_{\rm L}/d_{\rm L}\lesssim 0.3-0.5$), especially when the information on the mass and mass ratio provided by electromagnetic observations are considered.
Assuming that the redshifts of ``PTAs detected'' SMBBHs can be obtained from electromagnetic observations, the number of those SMBBHs, typically with chirp mass in the range from $10^9$ to $10^{10} M_{\odot}$ and frequency in the range from $10^{-8}$ to $10^{-9}$\,Hz, that can be used as standard sirens, is expected to be upto hundred thousands. The number of SMBBHs that is expected to be detected with SNR$>10$ and $>50$ by future SKA-PTA with conservative (or optimistic) settings are $\sim 200$ (or $\sim 3000$) and $\sim 60$ (or $\sim 1500$), respectively. Using these SMBBHs as standard sirens can put constraint on the EoS of dark energy to an uncertainty of $\Delta w \sim 0.1$ (or $\sim 0.02$). If the chirp mass for SMBBHs can be obtained by electromagnetic observations to a precision of $\sigma \ln M_{\rm c} \la 0.5$ or higher, the number of ``PTAs detected'' SMBBHs with SNR$>10$ and $>50$ can be $\sim 30,000-242,000$ and $\sim 1000-10,000$, respectively, depending on different PTAs settings. With these SMBBHs, the constraint on the EoS of dark energy can be $\Delta w \lesssim 0.01-0.1$.

In our simple model to generate SMBBHs by major mergers of galaxies, the produced SMBBHs are assumed to have the same mass ratio as those of their merging progenitor galaxies. In reality, those SMBBHs may have a mass ratio different from their parent merging galaxies, partly due to the scatters in the $M_{\bullet}-M_{gal}$ relation adopted to estimate black mass in individual galaxies and partly due to the growth of black holes during the merging processes. The former one may lead to more massive black holes, and the latter one may cause the increase of the mass ratio and thus the chirp mass of SMBBHs because the secondary black holes may accrete more gas than the primary ones. Therefore, the number of SMBBHs that can be used as the standard sirens may increase significantly, and consequently leads to better constraint on the EoS of dark energy presented above.

We also note that it is important to get accurate distance measurements in order to use the ``PTAs detected'' SMBBHs as standard sirens. However, most ``PTAs detected'' SMBBHs should have GW freqency $\lesssim 10^{-8}$\,Hz and negligible frequency change rates, which hinder accurate measurements of redshifted chirp mass and luminosity distance. On the one hand, if there are a number of SMBBHs (e.g., $100$), like those among the SMBBH candidates (bottom-left panel of Fig.~\ref{f8}, with $f_0\sim 10^{8}-10^{-7}$\,Hz) can have significant frequency change rate, their chirp masses and distances may be well determined from the GW signals itselves, and thus lead to constraints on the EoS of dark energy with considerable precision (e.g., $\Delta w \lesssim 0.1$). On the other hand, if the total mass and mass ratio of SMBBHs, and thus the chirp mass, can be obtained from electromagnetic monitoring of those systems with high accuracy, then the luminosity distance measurements would be improved a lot. However, this requires to do dedicated studies for many individual ``PTAs detected'' SMBBHs.

\begin{acknowledgments}
We appreciate the helpful discussions with Xingjiang Zhu, Yan Wang, Linqing Wen and kejia Lee.
This work is partly supported by the National Natural Science Foundation of China (Grant No. 11690024, 11773028, 11603020, 11633001, 11173021, 11322324, 11653002, 11421303, and 11873056), the National Key Program for Science and Technology Research and Development (Grant No. 2016YFA0400704), and the Strategic Priority Program of the Chinese Academy of Sciences (Grant No. XDB 23040100).
\end{acknowledgments}

\end{document}